\documentclass[aps,prd,twocolumn,showpacs,nofootinbib]{revtex4-1}

\usepackage{amssymb} \usepackage{amsmath} \usepackage{graphicx}
\usepackage{epsfig,latexsym}

\RequirePackage{xspace} \allowdisplaybreaks

\begin{document}

\def\bef{\begin{figure}}
\def\eef{\end{figure}}

\newcommand{\nl}{\nonumber\\}

\newcommand{\ans}{ansatz }
\newcommand{\be}[1]{\begin{equation}\label{#1}}
\newcommand{\beq}{\begin{equation}}
\newcommand{\ee}{\end{equation}}
\newcommand{\beqn}[1]{\begin{eqnarray}\label{#1}}
\newcommand{\eeqn}{\end{eqnarray}}
\newcommand{\bd}{\begin{displaymath}}
\newcommand{\ed}{\end{displaymath}}
\newcommand{\mat}[4]{\left(\begin{array}{cc}{#1}&{#2}\\{#3}&{#4}
\end{array}\right)}
\newcommand{\matr}[9]{\left(\begin{array}{ccc}{#1}&{#2}&{#3}\\
{#4}&{#5}&{#6}\\{#7}&{#8}&{#9}\end{array}\right)}
\newcommand{\matrr}[6]{\left(\begin{array}{cc}{#1}&{#2}\\
{#3}&{#4}\\{#5}&{#6}\end{array}\right)}
\newcommand{\cvb}[3]{#1^{#2}_{#3}}
\def\lsim{\raise0.3ex\hbox{$\;<$\kern-0.75em\raise-1.1ex
e\hbox{$\sim\;$}}}
\def\gsim{\raise0.3ex\hbox{$\;>$\kern-0.75em\raise-1.1ex
\hbox{$\sim\;$}}}
\def\abs#1{\left| #1\right|}
\def\simlt{\mathrel{\lower2.5pt\vbox{\lineskip=0pt\baselineskip=0pt
           \hbox{$<$}\hbox{$\sim$}}}}
\def\simgt{\mathrel{\lower2.5pt\vbox{\lineskip=0pt\baselineskip=0pt
           \hbox{$>$}\hbox{$\sim$}}}}
\def\unity{{\hbox{1\kern-.8mm l}}}
\newcommand{\eps}{\varepsilon}
\def\ep{\epsilon}
\def\ga{\gamma}
\def\Ga{\Gamma}
\def\om{\omega}
\def\omp{{\omega^\prime}}
\def\Om{\Omega}
\def\la{\lambda}
\def\La{\Lambda}
\def\al{\alpha}
\newcommand{\ov}{\overline}
\renewcommand{\to}{\rightarrow}
\renewcommand{\vec}[1]{\mathbf{#1}}
\newcommand{\vect}[1]{\mbox{\boldmath$#1$}}
\def\tm{{\widetilde{m}}}
\def\mcirc{{\stackrel{o}{m}}}
\newcommand{\Dm}{\Delta m}
\newcommand{\dm}{\varepsilon}
\newcommand{\tanb}{\tan\beta}
\newcommand{\nbar}{\tilde{n}}
\newcommand\PM[1]{\begin{pmatrix}#1\end{pmatrix}}
\newcommand{\up}{\uparrow}
\newcommand{\down}{\downarrow}
\def\omE{\omega_{\rm Ter}}
%

\newcommand{\Dsusy}{{susy \hspace{-9.4pt} \slash}\;}
\newcommand{\DCP}{{CP \hspace{-7.4pt} \slash}\;}
\newcommand{\mc}{\mathcal}
\newcommand{\gr}{\mathbf}
\renewcommand{\to}{\rightarrow}
\newcommand{\gtc}{\mathfrak}
\newcommand{\wh}{\widehat}
\newcommand{\br}{\langle}
\newcommand{\kt}{\rangle}


\def\lsim{\mathrel{\mathop  {\hbox{\lower0.5ex\hbox{$\sim$}
\kern-0.8em\lower-0.7ex\hbox{$<$}}}}}
\def\gsim{\mathrel{\mathop  {\hbox{\lower0.5ex\hbox{$\sim$}
\kern-0.8em\lower-0.7ex\hbox{$>$}}}}}

\def\nn{\\  \nonumber}
\def\de{\partial}
\def\brf{{\mathbf f}}
\def\bbf{\bar{\bf f}}
\def\bF{{\bf F}}
\def\bbF{\bar{\bf F}}
\def\bA{{\mathbf A}}
\def\bB{{\mathbf B}}
\def\bG{{\mathbf G}}
\def\bI{{\mathbf I}}
\def\bM{{\mathbf M}}
\def\bY{{\mathbf Y}}
\def\bX{{\mathbf X}}
\def\bS{{\mathbf S}}
\def\bb{{\mathbf b}}
\def\bh{{\mathbf h}}
\def\bg{{\mathbf g}}
\def\bla{{\mathbf \la}}
\def\bmu{\mathbf m }
\def\by{{\mathbf y}}
\def\bmu{\mbox{\boldmath $\mu$} }
\def\bsig{\mbox{\boldmath $\sigma$} }
\def\bunity{{\mathbf 1}}
\def\cA{{\cal A}}
\def\cB{{\cal B}}
\def\cC{{\cal C}}
\def\cD{{\cal D}}
\def\cF{{\cal F}}
\def\cG{{\cal G}}
\def\cH{{\cal H}}
\def\cI{{\cal I}}
\def\cL{{\cal L}}
\def\cN{{\cal N}}
\def\cM{{\cal M}}
\def\cO{{\cal O}}
\def\cR{{\cal R}}
\def\cS{{\cal S}}
\def\cT{{\cal T}}
\def\eV{{\rm eV}}
%

\title{Limiting First Order Phase Transitions in Dark Gauge Sectors \,\,\,\,\,\,\,\,\,\,\,\,\,\,\,\,\,\,\,\,\,\,\,\,\,\,\,\,\,\,\,\,\,\,\,\,\,\,\,\,\,\,\,\,\,\,\,\,\,\,\,\,\,\,\,\,\,\,\,\,\,\,\,\,\,\,\,\,\,\
from Gravitational Waves experiments  }

\author{Andrea Addazi$^1$}\email{andrea.addazi@infn.lngs.it}
\affiliation{$^1$ Dipartimento di Fisica,
 Universit\`a di L'Aquila, 67010 Coppito AQ and
LNGS, Laboratori Nazionali del Gran Sasso, 67010 Assergi AQ, Italy}

\begin{abstract}

We discuss the possibility to indirectly test First Order Phase Transitions 
of hidden sectors. We study the interesting example of a {\it dark
standard model} with a deformed parameter space in the Higgs potential.
A dark electroweak phase transition can 
be limited from next future experiments
 like eLISA and DECIGO.

\end{abstract}

\maketitle

\section{Introduction}

A new era of astrophysics and cosmology was opened with the 
detection of gravitational waves. 
For the first time, we have a chance to explore the dark 
side of the Universe from gravitational radiation. 
The possibility to test 
new physics
beyond the standard model of particles is particularly exciting.
For instance, it is conceivable that gravitational wave detectors 
could constrain or detect signals from unconventional candidates 
of dark matter beyond WIMP paradigm \footnote{
We mention that other interesting possibilities of GW emissions from collisions of bubbles
were recently suggested \cite{Schwaller:2015tja,Huang:2016odd,Artymowski:2016tme,Dev:2016feu,Katz:2016adq,Huang:2017laj}.}. 
For instance, if dark matter is composed by particles from a dark gauge sector, 
a dark first order phase transition in the early Universe will still be an open possibility. 
In this letter, we suggest to test 
first order phase transitions from 
hidden gauge sectors 
with gravitational wave detectors. 
In particular, we will focalize our analysis 
to the case of a Dark Standard Model: 
 $G_{SM}\times G'_{SM} \times Z_{2}$,
where $Z_{2}$ is a discrete symmetry guaranteeing that matter and gauge content, Yukawa and gauge couplings 
of the dark SM (D-SM)
are all equal to ordinary SM (O-SM) ones, i.e. $Z_{2}:G_{SM}\leftrightarrow G_{SM}'$. 
The idea of Dark Matter as a specular hidden standard model 
was largely explored in literature  
and it has many interesting consequences in astrophysics and cosmology, Dark Matter Direct Detection, Ultra Cold Neutrons and
Neutrino physics \cite{Blinnikov:1983gh,Blinnikov:1982eh,Khlopov:1989fj,Berezhiani:1995yi,Berezhiani:1995am,Berezhiani:1996sz,Bento:2001rc,Berezhiani:2005ek,Berezhiani:2011da,Berezhiani:2005hv,Addazi:2015cua,Berezhiani:2016ong,Foot:1991py,Foot:1993yp,Foot:1995pa,Foot:2008nw,Foot:2014xwa,Foot:2014osa}
\footnote{An alternative can be to introduce a dark strong sector with a low scale confinement, accounting for the correct abundance of 
dark matter and dark energy   \cite{Addazi:2016sot, Addazi:2016nok}.  
Another interesting alternative was discussed in Ref.\cite{Addazi:2016oob}, where a hidden Born-Infeld condensate could
generate a cosmological term as well as a neutrino mass and cold dark matter as a neutrino superfluid state. }.
The $Z_{2}$-symmetry can guarantee the specularity of O-SM and D-SM only 
at perturbative level. However, the Higgs potentials could receive extra non-perturbative 
self-interaction corrections from various unspecified dynamical mechanisms.
So that, the D-SM Higgs, dub $H'$ can have the same perturbative terms of O-SM Higgs
$H$, as $\mu^{2}|H'|^{2}+\frac{\lambda}{4}{|H'|^{4}}$, plus extra non-perturbative terms 
like $|H'|^{6}/\Lambda^{2}$. These extra terms can be particularly interesting in our case
because they can lower the double well barrier. 
As a consequence, during the dark electroweak phase transition at $T'\sim v\sim 200\, \rm GeV$,
the materialization of Bubbles is particularly efficient and their collisions 
generate detectable gravitational signals. 
For instance, this mechanism could explain the dark matter genesis
through a dark electroweak baryogenesis mechanism
\footnote{For complete reviews on electroweak baryogenesis models see Refs.\cite{Morrissey:2012db,Trodden:1998ym}.}. 
However, for a good satisfaction of Sakharov's criterions, 
an extra source of CP violating phases have to be introduced 
in the dark SM. 
For example, one can extend the minimal standard Model Higgs sector
with a Higgs doublet $H_{1,2}$ and consequently $H'_{1,2}$. 
In this framework a CP violation in the D-SM can be introduced
with a phase field which changes of a finite $\Delta \theta' \neq 0$ during 
the transition from a false to the true vacuum. 
A Dark electroweak baryogenesis,
induced by D-SM particle scatterings on Bubbles 
with a CP-violating phase field localized on it, 
can be related to a production of 
gravitational wave signals from Bubble-Bubble collisions.

\begin{figure}[htb] \label{BARI}
\begin{center}
\caption{GW spectra $h^{2}\Omega_{GW}$ as function of GW frequency 
is displayed in scale log-log scale  $\left(\log_{10}(f[Hz]),\log_{10}\,(h^{2}\Omega_{GW}^{2})\right)$
for various 
non-perturbative scales $\Lambda=590,600,650\, \rm GeV$,
(conventionally) assuming $(\kappa_{1}+\kappa_{3})=1$, $(\kappa_{2}+\kappa_{3})=1$, $(\kappa_{2}+\kappa_{4})=1$.
In Green and Blue, we display the approximated expected experimental sensitivity of future GW interferometers
eLISA C1 
 \cite{Caprini:2015zlo} and U-DECIGO \cite{Kudoh:2005as} respectively.}
\includegraphics[scale=0.09]{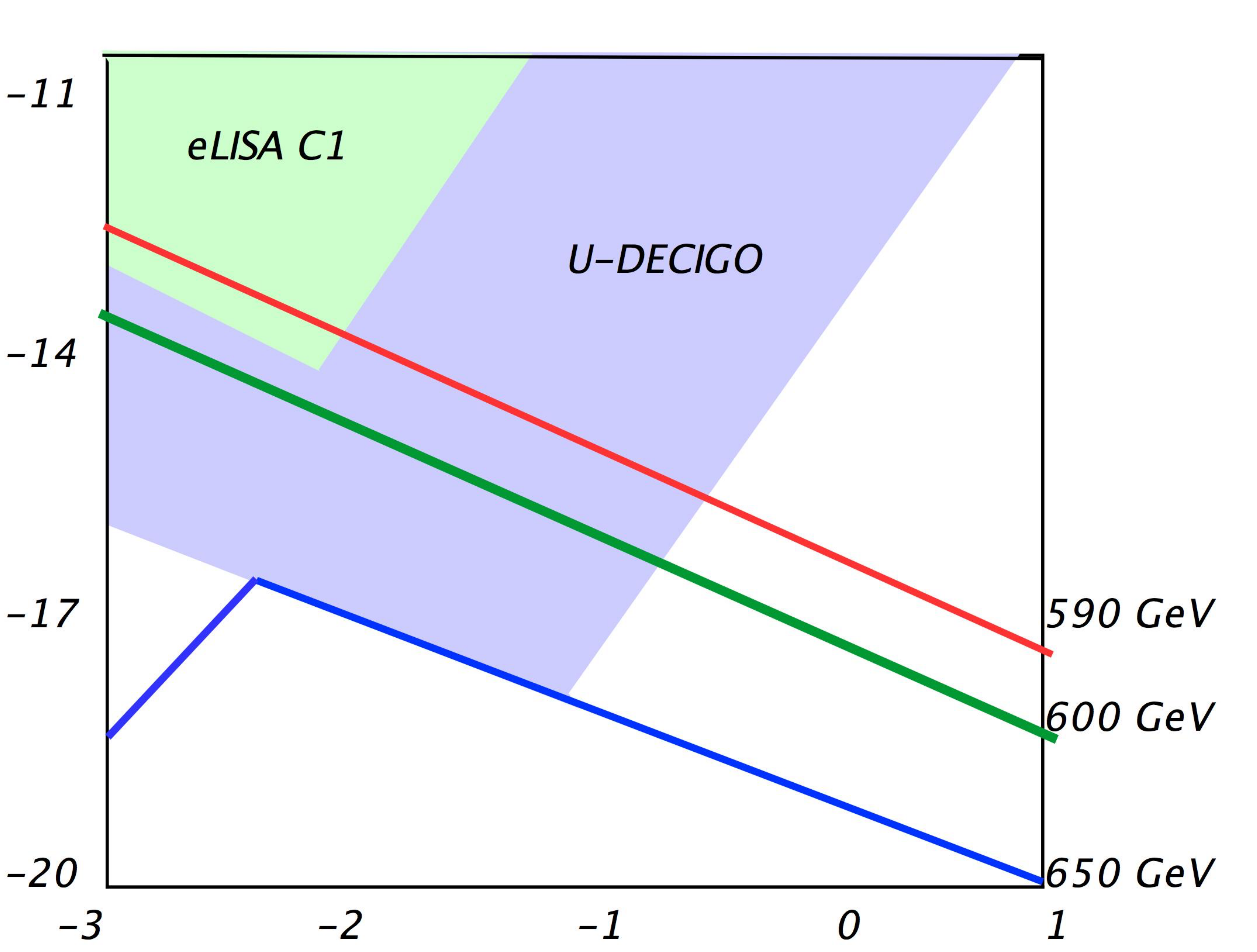}  
\end{center}
\vspace{-1mm}
\end{figure}

We will show that for $v\geq \Lambda \geq 1, \rm TeV$, the predicted signal can be detected 
by future experiments like eLISA (C1) and U-DECIGO. 
This letter is organized as follows:
In Section II we will introduce the full Higgs sector of our model, 
in Section III we will discuss an example of electroweak baryogenesis mechanism 
and gravitational signals from Bubbles collisions. 
In Section IV we show our conclusions and comments on possible extensions and UV completions of our model.

\section{Model}

The O- and D- Higgs sector has a general potential which reads as 
\be{Higgs}
V_{(0)}^{TOT}=V_{(0)}+V'_{(0)}+V^{mix}_{(0)}
\ee
$$V_{(0)}=V_{(0)}^{P}+V_{(0)}^{NP}$$
$$V_{(0)}'=V_{(0)}'^{P}+V_{(0)}'^{NP}$$
$$V^{mix}_{(0)}=V^{mix,P}_{(0)}+V^{mix,NP}_{(0)}$$
where perturbative terms are 
\be{VHH}
V_{(0)}^{P}=-\mu_{i}^{2}H_{i}^{\dagger}H_{i}+\lambda_{i}(H_{i}^{\dagger}H_{i})^{2}+\lambda (H_{i}H_{j})^{\dagger}(H_{i}H_{j})
\ee
\be{VHH2}
V_{(0)}'^{P}=-\mu_{i}^{2}H_{i}'^{\dagger}H'_{i}+\lambda_{i}(H_{i}'^{\dagger}H'_{i})^{2}+\lambda (H'_{i}H'_{j})^{\dagger}(H'_{i}H'_{j})
\ee
where $\lambda (H_{i}H_{j})^{\dagger}(H_{i}H_{j})$ and its mirror twin 
we mean to include $H_{1}^{\dagger}H_{1}H_{2}^{\dagger}H_{2}$,
$H_{1}^{\dagger}H_{2}H_{1}^{\dagger}H_{2}$ 
and $h.c$;
\be{suppr}
V^{mix,P}_{(0)}=-\bar{\kappa}(H_{1}H_{2})^{\dagger}(H'_{1}H'_{2})+h.c.
\ee
We assume that the only relevant non perturbative terms are D-ones:
\be{V6NP0}
V'^{NP}_{(0)}=\frac{\kappa_{1}}{8\Lambda^{2}}(H'_{i}H_{i}'^{\dagger})^{3}+\frac{\kappa_{2}}{8\Lambda^{2}}(H_{i}'H_{j}')^{3}
\ee
$$+\frac{\kappa_{3}}{8\Lambda^{2}}H_{i}'H_{i}'^{\dagger}(H'_{j}H_{j}'^{\dagger})^{2}+\frac{\kappa_{4}}{8\Lambda^{2}}(H'_{i}H_{i}'^{\dagger})(H'_{i}H_{j}')^{2}+h.c.$$
with $i\neq j$ and $i,j=1,2$,
where $\bar{\kappa}<10^{-8}$ in order to avoid a dangerous thermalization 
among the O- and D- SM, 
where $\kappa_{1,..,4}$  parametrize the difference between the Non perturbative scale in Eq.(\ref{V6NP0})
(See Ref.\cite{Berezhiani:2005ek} for a complete discussion on reheating limits).

In thermal bath, the finite-temperature effective potential up to one-loop is
\be{cowp}
V'^{(eff)}=V'^{(0)}+V'^{(1)}(T'=0)+\Delta V'^{(1)}(T')
\ee
which for $V'^{(1)}(T'=0)$ is the Coleman-Weinberg potential
while $V'^{(1)}(T')$ are thermal corrections.
\be{effective}
V^{(1)}(T'=0)\simeq \frac{c_{i}}{2}H'^{\dagger}_{i}H'_{i}
\ee
where 
\be{cD}
c_{1}=+c_{H_{1}}+\frac{1}{4}y_{t}^{2}+\frac{1}{16}(g^{2}_{Y}+3g^{2}_{2})
\ee
\be{cD}
c_{2}=+c_{H_{2}}+\frac{1}{4}y_{b}^{2}+\frac{1}{12}y_{\tau}^{2}+\frac{1}{16}(g^{2}_{Y}+3g^{2}_{2})
\ee
$$c_{H_{i}}=c_{H_{i}}^{P}-\frac{3}{4}c_{H_{i}}^{NP}$$
$$c_{H_{i}}^{P}=\frac{1}{2}\lambda_{i}+\frac{1}{4}\lambda+\frac{1}{12}\lambda_{S}$$
$$c_{H_{1}}^{NP}=\frac{1}{\Lambda^{2}}\left((\kappa_{1}+\kappa_{3})v_{1}^{2}+(\kappa_{3}+\kappa_{4})v_{2}^{2} \right)$$
$$c_{H_{2}}^{NP}=\frac{1}{\Lambda^{2}}\left((\kappa_{2}+\kappa_{3})v_{2}^{2}+(\kappa_{3}+\kappa_{4})v_{1}^{2} \right)$$
while thermal field theory corrections are understood and taken into account in the following section. 

\section{Dark Electroweak Baryogenesis and Gravitational waves}

For the Dark Electroweak Baryogenesis, 
a double-well (or multiple) potential between a metastable false vacuum 
and the true (or lower) one is postulated. 
The quantum tunneling process, mediated by a Coleman-De Luccia-like instanton, 
sources the materialization of a true vacuum bubble. 
The quantum tunneling probability nearby the dark electroweak scale $v_{D}$
is $O(1)$. In other words, $\Gamma(T'^{*})\simeq H^{4}(T'^{*})$, where $T'^{*}$ is the critical phase transition temperature, 
$\Gamma(T')$ is the tunneling transition rate and $H$ is the Hubble parameter. 
The three-dimensional Euclidean action of the Bubble is 
\be{S3}
S_{3}=\int d^{3}r\left[\frac{1}{2}(\nabla H'_{i})^{2}+V_{eff}(H_{i}',T') \right]
\ee
The proprieties of bubbles are parametrized by 
\be{alpha}
\alpha=\frac{\epsilon(T_{*}')}{\rho_{rad}(T')}
\ee  
where 
\be{epsilonTs}
\epsilon(T')=\left[T'\frac{dV_{eff}^{min}}{dT'}-V_{eff}^{min}(T')\right]_{T'=T_{*}'}
\ee
$\alpha$ is the ratio of the false vacuum energy density on the plasma thermal energy density
\be{rho}
\rho_{rad}^{D}(T')=\frac{\pi^{2}}{30}g_{*}(T')T'^{4}
\ee
In the Dark Electroweak Baryogenesis, the CP and B-violations are induced 
from scatterings of SM particles on bubbles. 
The O- and D- Higgs minima are
\be{HH12}
H_{i}=\frac{1}{\sqrt{2}}\left(0,\frac{v_{i}}{\sqrt{2}}e^{i\theta_{i}}\right),\,\,\,\,H_{i}'=\frac{1}{\sqrt{2}}\left(0,\frac{v_{i}'}{\sqrt{2}}e^{i\theta'_{i}}\right)
\ee
If $\theta_{1}'\neq \theta_{2}'$, CP-violating phases are introduced in the Yukawa couplings 
as $\theta'=\theta_{1}'-\theta_{2}'$. In particular, in couplings with quarks introduce a CP-violating 
phase in the CKM matrix as
\be{Yukawa}
\mathcal{L}_{U',D',H'_{1,2}}\rightarrow v_{1}Y_{U,gg'}e^{i\theta'_{1}}\bar{Q}_{L,g}'U_{R.g'}'+Y_{D,gg'}v_{2}e^{i\theta'_{2}}\bar{Q}_{L,g}'D_{R,g'}'+h.c.
\ee
where $g,g'$ are quark generations. 

In the limit of $\bar{\kappa}\rightarrow 0$ (O- and D-Higgs decoupling), 
the
Lagrangian for a Dark fermion $\psi'$ in the background of
a bubble wall with a CP-odd $\theta'$ field localized
on it
\be{CPV}
\mathcal{L}=i\bar{\psi'}\left(\partial_{\mu}+im_{\psi'}+\frac{i}{2}\frac{v_{2}^{2}}{v^{2}}\gamma^{5}\partial_{\mu}\theta' \right)\psi'
\ee
which can be obtained eliminating the CP violating phase $\theta_{1}$ in Eq.(\ref{Yukawa}) and after a hypercharge rotation transformation. 
Eq.(\ref{CPV}) can mediate a non-local electroweak baryogenesis mechanism \cite{Joyce:1994zn}.
In fact,  the Bubble wall propagating along a direction z with a velocity $U$
\be{psimu}
\mu_{\psi'}=\frac{U\log 2}{3\zeta(3)}\frac{v_{2}^{2}}{v^{2}}\partial_{z}\theta\left(\frac{m_{\psi'}}{T'}\right)^{2}
\ee
(where $\zeta$ is the Riemann function)
 sources the Boltzaman equation 
\be{nB}
\frac{dn_{B}}{dt}=-\frac{(\#_{f})\Gamma(T')}{2T'}\sum_{i}\mu_{i}
\ee
where $\#_{f}$ is the number of families and $\mu_{i}$ are chemical potentials of 
LH fermions of species $i$. 
The Dark Baryon asymmetry can be calculated with the same methods
used in Ref.\cite{Joyce:1994zn}. 
We obtain the following result: 
\be{nBs}
B'=\frac{n_{B'}}{s'}=C\frac{\eta}{D}\frac{45}{4g'_{*}\pi^{4}}U\left(\frac{m_{f}}{T'}\right)^{2}\left(\frac{1}{2T'L_{W}}\right)\Delta \theta_{CP}'
\ee
where $g'^{*}=g_{*}(T')$,
where $C$ depends on various approximation regimes: 
$$C=-\frac{\Gamma_{s}D_{L}}{U^{2}},\,\,\,\,U^{2}>\Gamma_{\tau}D_{R},\Gamma_{s}D_{L}$$
$$C=+\frac{2}{3}\frac{\Gamma_{s}D_{R}}{U^{2}},\,\,\,\,\Gamma_{\tau}D_{R}>U^{2}>\Gamma_{s}D_{L}$$
$$C=+\frac{2}{3},\,\,\,\,\Gamma_{\tau}D_{R},\Gamma_{s}D_{R}>U^{2}$$
$$C=-\frac{1}{2}\frac{U}{\sqrt{\Gamma_{ss}D_{q}}}\frac{3\Gamma_{s}D_{q}}{U^{2}},\,\,\,\,\frac{\delta D_{q}}{D_{q}}<\sqrt{\frac{U}{\Gamma_{ss}D_{q}}}<1$$
$$C=-\frac{1}{2}\frac{\delta D_{q}}{D_{q}}\frac{3\Gamma_{s}D_{q}}{U^{2}},\,\,\,\,\frac{\delta D_{q}}{D_{q}}>1\,\,\,\,U^{2}>\Gamma_{s}D_{q}$$
where
$$D_{L}^{-1}=8\alpha_{W}^{2}(1+0.8 \tan^{4}\theta_{W})T'\simeq \frac{T'}{100}$$
$$D_{R}^{-1}\simeq 28\alpha_{W}^{2}\tan^{4}\theta_{W}\,T'\simeq \frac{T'}{380}$$
$$D_{q}^{-1}\simeq 8\alpha_{s}^{2}T'\simeq \frac{T'}{6}$$
$$\Gamma_{q}=0.2\alpha_{s}y_{t}^{2},\,\,\,\Gamma_{LR}=\Gamma_{\tau_{R}}=2\Gamma_{\tau_{L}}=0.3\alpha_{W}y_{\tau}^{2}$$
$$\Gamma_{s}=6N_{F}\kappa_{s}\alpha_{W}^{4}T'\simeq 2\times 10^{-5}\kappa_{s}T'$$
$$\Gamma_{ss}=64\kappa_{ss}\alpha_{S}^{4}T'\simeq \kappa_{ss}\frac{T'}{40}$$
where $L_{W}$ is the wall thickness, 
$D_{L,R,q}$ are diffusion coefficients of LH, RH particles and quarks, 
$\Gamma_{q,LR}$ perturbative decay rates in plasma
mediated by Higgs bosons,
$\kappa_{s,ss}$ are numerically estimated constants in the range $0.1\div 1$, 
 $y_{t,\tau}$ Yukawa couplings of top and $\tau$
 and
$\eta$ is the persistent length of the initially injected particles currents.

The Cold Dark Matter coincidence $\Omega_{CDM}\simeq 5 \Omega_{B}$
can be recovered from electroweak baryogenesis with the right choice of initial 
parameter conditions in Eq.(\ref{nBs}). 

An inevitable prediction of this scenario is that Bubbles inevitably will collide 
each other, producing gravitational waves. 
The frequency and intensity of the produced signal is controlled by the 
Dark SM couplings, Dark phase transition critical temperature,
number of dark degree of freedoms $g'^{*}$,
Bubbles velocity and Higgs potential shape. 

In particular, we can estimate frequency and intensity using 
similar calculations of Refs.\cite{Kamionkowski:1993fg,Delaunay:2007wb}
and more recently \cite{Jinno:2016vai}.
We obtain:
\be{naive}
f_{Collisions}\simeq 5\times c_{i}\left( \frac{\beta}{H_{*}}\right)\left(\frac{T'^{*}}{100\, \rm GeV} \right)(g_{*}')^{1/6}\,\rm Hz
\ee
\be{OmegaCo}
\Omega_{Collisions}h^{2}\simeq c_{i}\epsilon^{2}\left(\frac{H_{*}}{\beta} \right)^{2}\left(\frac{\alpha}{1+\alpha} \right)
\left(\frac{U^{3}}{0.24+U^{3}}\right)\left(\frac{100}{g'^{*}} \right)^{1/3}
\ee
$U$ bubble velocity. 
Let us note that the case $T'_{0}\geq T_{0}$ with $T_{0}$ temperature $\leq T_{BBN}$
 is already excluded by BBN and CMB constrains on sterile neutrini. 
In fact, 
\be{NNN}
\Delta N_{\nu}=6.14\, \left(\frac{T'_{0}}{T_{0}} \right)^{4}
\ee
and, in order to have $\Delta N_{\nu}<1$, $T'_{0}/T_{0}\leq 0.64$. 
So that, the dark first order transition has necessary to cool the dark sector
down to $T_{0}'<0.64\,T_{0}$. 
Choosing $T'\simeq 0.5 T$ 
asymmetric reheating mechanisms
and various values of $\Lambda=590,600,650\, \rm GeV$, we show 
interesting examples of gravitational wave spectrum predicted by 
the dark electroweak phase transition 
in Fig.1.
The discussion of a specific mechanism for a $T'\neq T$ is beyond the purposes of this letter. 
But it can be understood as an asymmetric reheating of the O- and D- sectors, 
i.e. the inflaton field is asymmetrically coupled to the two sectors.
Of course an exactly $Z_{2}$-symmetric lagrangian should guarantee 
democratic inflaton couplings with the two sectors.
However, non-perturbative corrections can generating soft breaking corrections 
to inflaton couplings with the two sectors. 
In conclusions, we will comment on possible UV origin of extra terms. 
Finally, we show a concrete example of dark baryon production compatible 
with sterile neutrino limits:
for $m_{f'}/T'\simeq y_{f}'=y_{f}$,
Eq.(\ref{nBs}) is
$g_{*}'U(y_{f}/L_{W}T')^{2}\Delta \theta_{CP}'$, 
where $L_{W}$ is the wall thickness;
taking $\eta=6D v$,
with $v=(1/4\ln 2)(1/2L_{W}T')$,
$U>2.1y_{\tau}$;
assuming $T'\sim 0.5 T$, 
we find 
$$\frac{n_{B'}}{s'}\simeq -\frac{8}{g_{*}'}\frac{y_{\tau}^{2}}{(L_{W}T')^{2}}\frac{\kappa_{s}}{U}\alpha_{W}^{2}\Delta \theta_{CP}'$$
with $L_{W}T'\sim 20,$ $g_{*}'\simeq 100$ and $U\simeq 0.1$
we obtain 
$$\frac{n_{B'}}{s'}\sim -2.4 \times 10^{-10}\kappa_{s}\Delta \theta_{CP}'$$
which approximately must saturate five times the BBN bound $(4\div 11)\times 10^{-11}$
in order to recover the coincidence $\Omega_{DM}\simeq 5\Omega_{B}$,
i.e. implying  large dark CP violation $\Delta \theta_{CP}'\sim 1$.

\section{Conclusions and remarks}  
  
In this letter, we have explored the possibility to detect gravitational wave signals from 
a dark electroweak baryogenesis (DEB) of a dark standard model. 
DEB is one of the simplest classes of mechanisms explaining
dark matter genesis from a dark standard model. 
It predicts the materialization of Bubbles and their collisions  generate
a characteristic gravitational spectrum detectable in the next generation of experiments
like eLISA (C1) and U-DECIGO. 
As a consequence, gravitational wave astronomy can provide 
an important test for dark standard model, which is a good candidate 
of dark matter. Future data from gravitational waves experiments
can provide important limits on dark sectors. 

We have discussed the example of a dark non-local electroweak baryogenesis
where CP-asymmetry introduced from a two Higgs model.
On the other hand, in principle, our model can be embedded 
in a supersymmetric 
$G_{SM}\times G'_{SM} \times Z_{2}$. In this 
case, SUSY introduces extra CP-violating phases affecting the Baryogenesis mechanism. 
Supersymmetry can be asymmetrically broken 
among the two sectors by non-perturbative effects. It is possible that supersymmetry 
is broken at the electroweak scale or even at smaller scales in the dark sector, providing 
extra CP violating phase, while in our sector broken at higher scales. 
For example it is conceivable that SUSY could be dynamical broken by instantonic effects at high scales in our ordinary sector 
while gravitational or gauge mediators from the O-SM to D-SM transmit the SUSY breaking information. 
This mechanism naturally guarantees a hierarchy $M_{SUSY}>>M'_{SUSY}$ where 
$M_{SUSY},M'_{SUSY}$ are susy breaking scales of O-DM and D-SM respectively. 
In this case, dark gravitinos or other SUSY particle can have a mass close upon the electroweak scale
and in principle they can decay injecting extra neutrons or photons during dark BBN,
completely changing the dark nuclei composition of the dark sector 
with respect to O-BBN \footnote{Limits on supersymmetric particles 
from BBN were firstly suggested in \cite{Khlopov:1984pf}.}.
On the other hand, our model can be UV completed in context intersecting D-brane parallel worlds.
In this case, extra non-perturbative terms like $H_{1}^{6},H_{2}^{6},...$ softly breaking the $Z_{2}$ symmetry 
can be generated by exotic E-brane instantons, similarly to mechanisms suggested in 
our recent papers in various contexts \cite{Addazi:2015yna,Addazi:2015ewa,Addazi:2016xuh,Addazi:2016mtn,Addazi:2016mtn} \footnote{We suggest that it could be interesting to explore possible 
effects of non-local interaction terms in electroweak baryogenesis in context of effective non-local quantum field theories
like ones studied in Refs.\cite{Addazi:2015ppa} and embedding the SM of particles. 
}.
As well, inflaton couplings with the O- and D- sectors are the same at perturbative level, 
while exotic instantons can generate soft $Z_{2}$-breaking corrections 
leading to an asymmetric reheating of the O- and D- sectors ($T'\neq T$). 
Let us also remark that a dark standard model scenario 
weakly interacting with our ordinary sector
is also strongly motivated by solutions of the hierarchy problem 
introducing N parallel sector and obtaining a low cutoff $\Lambda_{UV}\sim M_{Pl}/\sqrt{N}$
\cite{Dvali:2009ne,Arkani-Hamed:2016rle}.
We conclude remarking that 
searches for dark first order phase transitions
with gravitational wave detectors are strongly motivated by 
dark matter genesis mechanisms
such as dark electroweak baryogenesis 
in context of
dark standard model dark matter (DSDM).
This could be a new interesting paradigm 
in dark matter phenomenology from gravitational radiation.

\vspace{1cm} 

{\large \bf Acknowledgments} 
\vspace{3mm}

I would like to thank Zurab Berezhiani for several useful discussions and remarks on these subjects. 
My  work was supported in part by the MIUR research
grant "Theoretical Astroparticle Physics" PRIN 2012CPPYP7
and  by
SdC Progetto Speciale Multiasse La Societ\'a  della Conoscenza in
Abruzzo.

\end{document}